\documentclass[11pt,a4paper]{article}

\usepackage{jheppub}
\usepackage{epsfig,amssymb,amsmath,psfrag,caption,subcaption}

\usepackage[T1]{fontenc}
\usepackage{psfrag}

\usepackage{url}

\def\N{{\cal N}}

\catcode`\@=11
\def \be  {\begin{equation}}
\def \ee  {\end{equation}}
\def \ba  {\begin{eqnarray}}
\def \ea  {\end{eqnarray}}
\def \baa {\begin{eqnarray*}}
\def \eaa {\end{eqnarray*}}
\def \bb  {\begin {thebibliography} }
\def \eb  {\end{thebibliography}}

\def \beqa {\begin{eqnarray} }
\def \eeqa {\end{eqnarray}}

\begin{document}

\bibliographystyle{JHEP}

\subheader{DESY 14-076}

\title{Heptagon Amplitude in the Multi-Regge Regime}

\author[a]{J.~Bartels}
\author[b]{V.~Schomerus}
\author[b]{M.~Sprenger}

\emailAdd{joachim.bartels@desy.de}
\emailAdd{volker.schomerus@desy.de}
\emailAdd{martin.sprenger@desy.de}

\affiliation[a]{II. Institute Theoretical Physics,\\
Hamburg University, Germany}
\affiliation[b]{DESY Theory Group,\\
Hamburg, Germany}

\abstract{
As we have shown in previous work, the high energy limit of scattering
amplitudes in $\N=4$ supersymmetric Yang-Mills theory corresponds to the
infrared limit of the 1-dimensional quantum integrable system that solves
minimal area problems in $AdS_5$. This insight can be developed into a
systematic algorithm to compute the strong coupling limit of amplitudes
in the multi-Regge regime through the solution of auxiliary Bethe Ansatz
equations. We apply this procedure to compute the scattering amplitude
for $n=7$ external gluons in different multi-Regge regions at infinite
't Hooft coupling. Our formulas are remarkably consistent with the expected form of $7$-gluon Regge cut contributions in perturbative gauge theory. A full description of the general algorithm and a derivation of results will be given in a forthcoming paper.}

\keywords{AdS/CFT, gluon amplitudes, Regge limit, thermodynamic Bethe Ansatz}

\maketitle

\section{Introduction}

The solution of a 4-dimensional gauge theory remains a visionary
dream of theoretical physics. Over the last years much progress
has been made in the planar limit of $\N=4$ supersymmetric Yang-Mills
theory (SYM). New methods based on symmetries and unitarity cuts
have helped to re-organize perturbative computations e.g.\ of scattering
amplitudes and have allowed to push them to previously unattainable
orders, see for example \cite{amplitudes2013} for the most recent results in
this dynamical research area. Even with all this progress, however,
the way from high to all orders still seems long and difficult. On
the other hand, through the celebrated AdS/CFT correspondence,
strongly coupled $\N=4$ SYM theory is related to superstring theory
on $AdS_5 \times S^5$ \cite{Maldacena:1997re}. This duality severely
constrains the strong coupling behavior of gauge theory amplitudes.
Some of these constraints will be discussed below.

In gauge theories, scattering amplitudes in the high energy regime are
of particular relevance. Remarkably, this regime is also computationally
more accessible than generic kinematics. In fact, it has been known for
a long time that integrable Heisenberg spin chains enter the expressions
for scattering amplitudes in the so-called multi-Regge (high energy)
limit \cite{Lipatov:1993yb,Faddeev:1994zg}. To leading logarithmic order,
this is even true for QCD. Given these features of the multi-Regge
limit in gauge theory one may
wonder about the nature of the corresponding limit for string theory
on $AdS_5$. In a previous paper we showed that the high energy limit for
scattering amplitudes in $\N=4$ SYM corresponds to the infrared (low
energy) limit of a 1-dimensional quantum integrable system for strings
on $AdS_5$ \cite{Bartels:2012gq}.

At strong coupling, scattering amplitudes in $\N=4$ SYM are believed
to coincide with the area of a minimal 2-dimensional surface that
approaches the boundary of $AdS_5$ along a light-like polygon
\cite{Alday:2007hr}. The
latter encodes all the kinematic data of the process. This minimal
area problem was shown to possess an intriguing reformulation in
which the area is reproduced by the free energy of a 1-dimensional
quantum integrable system \cite{Alday:2009dv,Alday:2010vh}. The
particle content and interactions
of the latter are designed so as to solve the original geometric
minimal area problem. The 1-dimensional quantum system contains
a number of mass parameters and chemical potentials which matches
precisely the number of kinematic invariants in the scattering
process. For generic kinematics and particle densities of the
1-dimensional system one must solve a complicated system of coupled
non-linear integral equations. These get replaced by a set of
algebraic Bethe Ansatz equations in the infrared limit of the
1-dimensional system, i.e.\ when we enter the multi-Regge regime
of gauge theory, see \cite{Bartels:2014} for details. 

The relations we have described in the previous paragraph allow
us to find exact analytic results for scattering amplitudes
in the multi-Regge regime of strongly coupled $\N=4$ SYM theory.
Below we shall provide explicit expressions for $n=7$ external
gluons. Let us recall that scattering amplitudes for $n \leq 5$
external gluons are entirely fixed by dual conformal symmetry.
A formula for $n=6$ was given recently in \cite{Bartels:2013dja},
see also \cite{Bartels:2010ej}. The extensions to $n=7$ we shall
describe below are new. Their detailed derivation will be given
in our forthcoming paper \cite{Bartels:2014}.

The formulas we shall spell out below turn out to be very nicely consistent 
with the results of perturbative gauge theory at weak coupling. The latter 
will be discussed in section 3 before we present our expressions for the 
scattering amplitudes with $n=7$ external gluons in section 4. Section 2 
contains some background information of kinematical invariants and the
multi-Regge limit.

\section{Kinematics}

In this article we consider $2 \rightarrow n-2$ production amplitudes
for massless external gluons. Our conventions for the enumeration
of momenta is shown in figure 1. Given the $n$ momenta $p_1, \dots,
p_n$ satisfying $p_i^2 = 0$ we can build the following Mandelstam
invariants
\begin{equation} \label{eq:xij2}
x^2_{ij}=\left(p_{i+1}+\cdots+p_j\right)^2,
\end{equation}
where $i,j = 1, \dots, n$. These invariants are related by momentum conservation and Gram determinant
relations so that only $3n-10$ of them are actually independent.
\begin{figure}[htb]
\centering
\epsfig{file=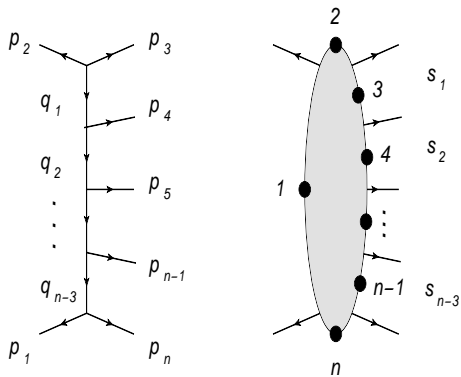, width=6cm,height=5cm}
\caption{Kinematics of the scattering process $2 \to n-2$. On the right-hand side we show a graphical representation of the dual variables $x_i$.}
\label{fig:configuration}
\end{figure}
Our description of scattering amplitudes below will be through the
so-called remainder function $R_n$. The latter depends on the
kinematic invariants only through $3n-15$ dual conformally invariant
cross ratios. For the discussion of the multi-Regge limit, the
following choice is very useful:
\begin{align}
u_{1\sigma}&=\frac{x^2_{\sigma+1,\sigma+5}x^2_{\sigma+2,\sigma+4}}
{x^2_{\sigma+2,\sigma+5}x^2_{\sigma+1,\sigma+4}},\label{eq:cr1}\\
u_{2\sigma}&=\frac{x^2_{\sigma+3,n}x^2_{1,\sigma+2}}
{x^2_{\sigma+2,n}x^2_{1,\sigma+3}},\label{eq:cr2}\\
u_{3\sigma}&=\frac{x^2_{2,\sigma+3}x^2_{1,\sigma+4}}
{x^2_{2,\sigma+4}x^2_{1,\sigma+3}},
\label{eq:cr3}
\end{align}
where $\sigma=1,...,n-5$. In the case of $n=6$ external gluons,
there are only 3 independent cross ratios, which we shall simply
denote by $u_1,u_2,u_3$. The number doubles when we go to $n=7$.

In the multi-Regge limit, the cross ratios $u_{1\sigma}$ tend to
$u_{1\sigma} \sim 1$ while the remaining ones tend to zero,
i.e.\ $u_{2\sigma},u_{3\sigma} \sim 0$. Cross ratios with
the same index $\sigma$ approach their limiting values such that the
following reduced cross ratios remain finite
\begin{equation}
  \label{eq:crsinmrl}
\left[\frac{u_{2\sigma}}{1-u_{1\sigma}}\right]^{\text{MRL}}
\hspace*{-6mm} =:
\frac{1}{|1+w_\sigma|^2}\quad , \quad
\left[\frac{u_{3\sigma}}{1-u_{1\sigma}}\right]^{\text{MRL}}
\hspace*{-6mm}=:
 \frac{|w_\sigma|^2}{|1+w_\sigma|^2}\, .
\end{equation}
Through these equations we have introduced the $n-5$ complex
parameters $w_\sigma$.
\begin{figure}[htb]
\centering
\epsfig{file=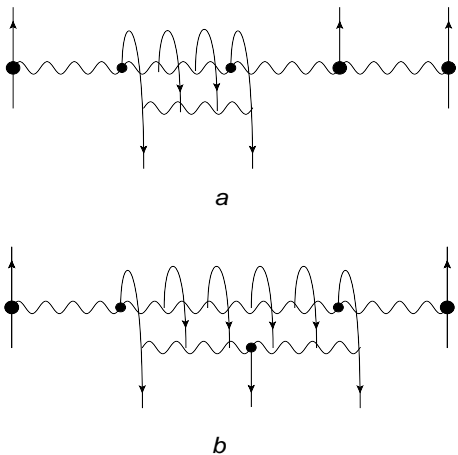, width=6cm,height=6cm}
\caption{Graphical representation of two Regge regions of the $7$-point amplitude.}
\label{fig:paths}
\end{figure}

An important aspect of the multi-Regge limit is the region in
which it is actually performed. We consider $2^{n-4}$ different
regions which depend on the sign of the energies of the produced particles $p^0_i$ for $i=
4, \dots, n-1$. Different regions can be reached from the one
in which all $p_i^0$ are positive by analytic continuation.
For $n=7$, there are eight different regions which we label
by $(\nu_1\nu_2\nu_3)$ with $\nu_i=${\it sgn}$(p_{i+3})$. The
region $(+++)$ is the original one in which all the $p^0_i$ are
positive. In this region, the multi-Regge limit of the remainder
function vanishes \cite{Bartels:2008ce}. The same is actually true for the three
regions $(-++),(+-+)$ and $(++-)$, which we therefore neglect in the following discussion. For the remaining four regions, things are
more interesting. Two of these regions are shown in figure 2. One of the missing regions in figure 2, $(+--)$, is related to $(--+)$ by a reflection along the
vertical axis. For the strong coupling analysis we have to define paths of analytic continuation to connect the different regions.
We choose to continue the $6$ basic cross ratios along the following curves $P_{\nu_1\nu_2\nu_3}$:
\begin{equation}
 \label{patha}
 \begin{array}{rlrl}
P_{--+}: u_{11}(\varphi) & =  e^{-2i\varphi} u_{11} & \ ,\
 u_{12}(\varphi) & =  u_{12}\ ,    \\[2mm]
 u_{21}(\varphi) & =   u_{21}   & \ ,\
u_{22}(\varphi)  & =   e^{-i\varphi} u_{22}\ , \\[2mm]
u_{31}(\varphi) & =   u_{31} & \ ,\
u_{32}(\varphi) & =  e^{i\varphi}u_{32}\ , \\[2mm]
\end{array} \end{equation}
\begin{equation}
 \label{pathc}
 \begin{array}{rlll}
P_{---}: u_{1\sigma}(\varphi) & =  e^{2i\varphi}
& ( 1-  \sqrt{1-e^{-2i\varphi}} &
\hspace*{-2mm}) u_{1\sigma}  \ ,   \\[2mm]
u_{21}(\varphi) & =   u_{21}   & \  ,\
u_{22}(\varphi)   =   u_{22} \ , & \\[2mm]
u_{31}(\varphi)  & =   u_{31} & \  ,\
u_{32}(\varphi)  =   u_{32} \ ,  & \\[2mm]
\end{array} \end{equation}
and
\begin{equation}
\label{pathb}
\begin{array}{rlrl}
P_{-+-}: u_{11}(\varphi) & =  e^{2i\varphi} u_{11} & \ ,\
 u_{12}(\varphi) & =  e^{2i\varphi} u_{12}\ , \\[2mm]
 u_{21}(\varphi) & =   e^{-i \varphi}  u_{21} & \ ,\
u_{22}(\varphi)  & =   e^{i\varphi} u_{22}\ , \\[2mm]
u_{31}(\varphi) & =  e^{i \varphi} u_{31} & \ ,\
u_{32}(\varphi) & =  e^{-i\varphi}u_{32}\ . \\[2mm]
\end{array} \end{equation}
Here, $\varphi \in [0,\pi]$ so that the cross ratios $u_{1\sigma}
(\varphi)$ come back to their starting values $u_{1\sigma}$ at the
endpoint of the path. The path $P_{+--}$ is obtained from $P_{--+}$
by the replacement $u_{11} \leftrightarrow u_{12}$ and $u_{2a}\leftrightarrow u_{3n-4-a}$. These paths and their construction will
be discussed in much more detail in our upcoming publication \cite{Bartels:2014}.

\section{Weak coupling}

We are now ready to discuss results on the multi-Regge limit
of scattering amplitudes for $n=7$ gluons in perturbative gauge
theory. Let us recall that in planar $\N =4$ super Yang-Mills
theory the full color-ordered maximally helicity violating
amplitude takes the following form
$$
A_n \sim A_n^{(0)} e^{F_n^{\text{BDS}} + R_n}\ .
$$
Here, $A_n^{(0)}$ is the tree level factor. The function
$F_n^{\text BDS}$ was introduced by Bern, Dixon and Smirnov
in \cite{Bern:2005iz}. It contains the known singular terms of the
amplitudes along with a relatively simple finite term. It
captures the amplitude correctly at one loop but misses
important contributions starting from two loops. The so-called
remainder function $R_n$ is IR-finite and invariant under dual
conformal transformations, i.e.\ it is a function of the
$3n-15$ cross ratios introduced above. While $R_n$ vanishes
for $n=4,5$, it must be non-zero starting from $n=6$ in order
to correct the unphysical analytical structure of the BDS
Ansatz, see \cite{Bartels:2008ce}.

The full remainder function $R_n$ was bootstrapped for $n=6$ up
to four loops \cite{Goncharov:2010jf,Dixon:2011pw,Dixon:2013eka,
Dixon:2014voa} and its symbol is known for $n > 6$ at two loops
\cite{CaronHuot:2011ky}. There exist further results for very
special kinematics, but these are not relevant for our
discussion. The most important results concern the multi-Regge
regime in which the cross ratios tend to the limit values
described above. Since the
remainder function contains branch cuts, its limiting behavior
depends on the multi-Regge region we consider. In fact, if we
perform this limit in the region where all $p_i^0 > 0$, then
the limit of the remainder function is trivial. In order to
take the limit in other regions, we must continue the cross
ratios. For $n=7$ gluons, the
regions of interest are $(--+)$, $(---)$ and $(-+-)$.
In all three cases the limiting behavior of the remainder
function $R_7$ is described by a factorizing ansatz. For
the region $(--+)$ this takes the form
\beqa
\label{R7--+}
\left[ e^{R_7+i\pi \delta_{7\, --+}}\right]^{\text{MRL}}_{--+} & = & \;i\frac{\lambda}{2} \sum_{n_1=-\infty}^{\infty} (-1)^{n_1} \int_{-\infty}^{\infty} \frac{d\nu_1}{2\pi}
| \Phi(\nu_1,n_1)|^2 \\[2mm]
& & \hspace*{2cm} \times \left| w_{1} \right|^{2i\nu_1} e^{-i n_1\phi_1}
\left((1-u_{11}) \frac{|w_1|}{|1+w_1|^2} \right)^{-\omega(\nu_1,n_1)}+...
\nonumber
\eeqa
Here, the phase $\delta_7$ denotes contributions from the analytic
continuation of $F_7^{\text{BDS}}$, the impact factor $\Phi$ is given by
an expansion in powers of the coupling (known up to N$^3$LO), and the
BFKL eigenvalues $\omega(\nu,n)$ are the lowest eigenvalues of
a non-compact SL(2,$\mathbb{C}$) Heisenberg Hamiltonian for a spin chain
of length two, see \cite{Lipatov:2009nt}. The parameters $\nu,n$ label
irreducible representations of SL(2,$\mathbb{C}$). Explicit expression
for $\omega(\nu,n)$ may be found in \cite{Bartels:2008sc,Fadin:2011we, Dixon:2014voa}.
Finally, the dots indicate that in addition to the Regge cut contribution there is still
a conformally invariant Regge pole term which, in the present context, is not of interest.
      
One may expect that the corresponding result for the second region is
a bit more complicated. This is indeed the case. For the region $(---)$ one
finds,
\beqa
\label{R7---}
\left[ e^{R_7+i\pi \delta_{7\, ---}}\right]^{\text{MRL}}_{---} & = &
i\frac{\lambda}{2} \sum_{n_1,n_2} (-1)^{n_1+n_2}\int\int \frac{d\nu_1 d\nu_2}{(2\pi)^2}
\  \Phi(\nu_1,n_1)^*   C(\nu_1,n_1;\nu_2,n_2) \Phi(\nu_2,n_2)\nonumber\\[2mm]
& & \times \prod_{\sigma = 1}^2 |w_\sigma|^{2i\nu_\sigma} e^{-i n_\sigma\phi_\sigma}
\left((1-u_{1\sigma}) \frac{|w_\sigma|}{|1+w_\sigma|^2}\right)^{-\omega(\nu_\sigma,
n_\sigma)} + \dots
\eeqa
The impact factors $\Phi$ and BFKL eigenvalues $\omega$ are the same as in eq.\eqref{R7--+}. The new production vertex $C$ has been calculated to LO in eqs.(19), (20) of \cite{Bartels:2011ge} . The leading order momentum space expression for the production vertex is real-valued, whereas the next-to leading order correction is expected to acquire a
phase. Eq.(\ref{R7---}) represents the leading contribution of this path, but is not
the full weak coupling result (as indicated by the dots). Namely, in addition to the
given integral, the remainder function contains a conformal Regge pole contribution
and two subtraction terms. In addition, we have to subtract the one loop terms from
the two energy factors in the second line of eq.\eqref{R7---}. The complete weak coupling result is given in \cite{Bartels:2013jna,BKL}. While these terms are relevant
for any comparison with a loop expansion of the 7-point amplitudes, in our discussion
it is sufficient to use eq.(\ref{R7---}).

Finally, for the region $(-+-)$, the leading order result is of the same form as eq.(\ref{R7---}): Differences between the two regions are expected to appear at NLO where the production vertex becomes complex-valued. For the region $(-+-)$ the production vertex is expected to be the complex conjugate of the one for the previous region $(---)$. Also, the subtraction terms (indicated in eq.(\ref{R7---}) by the dots) are different from those of the region  $(---)$.

It is important to note that, apart from the production vertex $C$, these expressions for the 7-point remainder functions are determined by functions which appear already in the 6-point remainder functions, in particular the
eigenvalue function $\omega(\nu,n)$.  This does not mean that the $n=6$ gluon results determine the multi-Regge limit of the remainder functions $R_n$ for $n \geq 7$.
In fact, already at $n=8$ gluons there exist kinematic regions in
which a new function appears. This is related to the fact that
the continuation along certain paths starts to probe eigenvalues
$\omega_3(\nu_1,n_1,\nu_2,n_2)$ of an open Heisenberg spin chain with three
sites. One of the remarkable outcomes of our analysis is that the close
relation between 6- and 7-point high energy amplitudes is also present at
strong coupling.

\section{Strong coupling}

Let us now come to the main new results of this work, namely the
computation of remainder functions in the multi-Regge regime of
strongly coupled $\N=4$ SYM theory. In particular, we shall spell
out formulas for the remainder function with $n=7$ external gluons.

As we briefly reviewed in the introduction, in strongly coupled
$\N=4$ SYM theory scattering may be reformulated as a minimal area
problem. The latter possesses an intriguing mathematical reformulation
through a 1-dimensional quantum integrable system which is composed of
particles whose interaction is fully characterized through $2\mapsto 2$
scattering phases. This system possesses $3n-15$ external parameters
which can be thought of as masses and chemical potentials. Given any
such system one needs to determine self-consistently the rapidity
densities of the various particles as a function of the external
parameters. This is done by solving a system of coupled non-linear
integral equations which involve both the external parameters and
the scattering phases. Roughly, such equations take the form
\begin{equation} \log Y(\theta) = - m \cosh \theta + C +
  \int\limits_{-\infty}^\infty d\theta'
  K(\theta-\theta') \log (1 + Y(\theta')) \ .
  \label{NLIE} \end{equation}
Here we have suppressed all the indices that run over the various
particles for simplicity. The parameters $m$ and $C$ play the role of the mass
parameter and chemical potential, respectively, and
\begin{equation} -2\pi i K(\theta) =: \partial_\theta \log S(\theta)
\label{KS}
\end{equation}
is associated with the scattering phase $S(\theta)$. Once the densities
$Y$ have been found, they can be used to determine the total energy of
the system.

There are two rather important observations concerning non-linear
integral equations of the form \eqref{NLIE}. The first one addresses
their behavior under analytic continuation of the parameters $m$ and
$C$. As we move the parameters through the complex plane, the solution $Y(\theta)$ to the integral equations changes. In particular, the solutions of the equation
$Y(\theta_*) = -1$ will wander through the space of complex rapidities
$\theta$. Looking at eq.\eqref{NLIE}
we can see that solutions of $Y(\theta_*) =-1$ are associated with
poles in the integrand of the non-linear integral equation. When these
poles cross the integration contour, the equation picks up a
residue contribution and hence assumes the new form
\begin{equation} \label{NLIES}
 \log Y(\theta) = -m'\cosh \theta  + C' + \sum_i \sigma_i
\log S(\theta-\theta_i) + \int\limits_{-\infty}^\infty d\theta'
  K(\theta-\theta') \log (1 + Y(\theta'))
\end{equation}
with sign factors $\sigma_i$ depending on whether the solution $\theta_i$
of $Y(\theta_i) =-1$ crosses from the lower half of the complex plane into
the upper half or vice versa. Whenever such a crossing takes place, a new contribution to the total energy of the system appears. One may interpret
these changes to the system as excitations that have been produced while
we continued the system parameters $m$ and $C$ to the values $m'$ and $C'$.

The second important observation concerning eq.\eqref{NLIE} addresses
its behavior in the limit of large mass parameter $m$. To be more precise,
the parameter $m$ is dimensionless and should rather be thought of as a product
$m = ML$ of a physical mass $M$ and the system size $L$. Sending $m$ to
infinity is then achieved be making the physical mass $M$ large or by going
to the limit of large system size. In such a limit, the first term on the
right-hand side \eqref{NLIE} goes to minus infinity and hence the function
$Y(\theta)$ that appears in the logarithm on the left-hand side must
approach zero. This in turn implies that $\log(1 + Y (\theta')) \sim 0$.
Thus we can neglect the integral on the right-hand side of our
non-linear integral equations. We are therefore left with the first two
terms on the right-hand side hand side of eq.\eqref{NLIES}. If we evaluate
this equation at the points $\theta = \theta_j$, use that $Y(\theta_j) = -1$
and exponentiate we obtain a set of algebraic Bethe Ansatz equations,
$$ e^{m'\cosh\theta_j-C'} = \prod_{i\neq j} S^{\sigma_i}(\theta_j-\theta_i). $$
These can be solved for the rapidities $\theta_j$ of the excitations.
Once they have been determined we can compute the energy as a sum of
energies, one for each crossed solution of $Y(\theta_i) = -1$.

All this applies to the minimal area problem in $AdS_5$. In that case
there exist $n-5$ complex mass parameters $m_s$ and $n-5$ purely complex chemical potential $C_s$. When the $m_s$ are sent to infinity in a certain
direction of the complex plane such that
\begin{equation} \label{limit}
 m_s e^{i(s-1)\frac{\pi}{4}} \rightarrow  \infty + i \log v_s
\end{equation}
with both $C_s$ and $v_s$ fixed and real, then the cross ratios of the
process approach the multi-Regge regime \cite{Bartels:2012gq}. The limit
is parametrized by $n-5$ complex variables $w_\sigma = w_\sigma(v_s,C_s)$ (cf.\ eq.(\ref{eq:crsinmrl}))
that can be scanned by adjusting $v_s$ and $C_s$.

In the large volume limit \eqref{limit}, the energy of the
1-dimensional quantum system vanishes and this in turn implies that the
remainder function vanishes in multi-Regge kinematics of strongly coupled
$\N=4$ SYM theory, as long as we are in the region with all $p_i^0$
positive. This is the same behavior as in the weakly coupled theory. In
order to obtain a non-vanishing result, we need to continue into another
multi-Regge region, as we discussed in the previous sections. This analytic continuation has to be performed before neglecting the integrals in eq.(\ref{NLIE}) since poles can cross the integration contour during the continuation, as explained before. We have worked
this out for a number of examples. Since the system parameters of the
1-dimensional quantum system are related to the cross ratios, we must
continue $m_s$ and $C_s$ along appropriate paths.

\begin{figure}
\begin{subfigure}[b]{.5\linewidth} \centering \epsfig{file=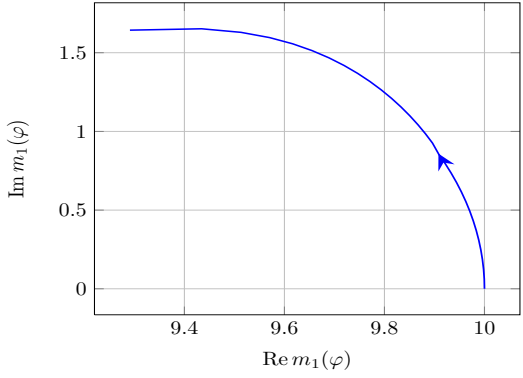,width=7cm,height=5cm}
\captionsetup{width=7cm}
\caption{Path of the system parameter $m_1$. Note that the
absolute value of $m_1$ remains large throughout the continuation.\\}
\label{fig:mmpm1}\centering
\end{subfigure}%
\begin{subfigure}[b]{.5 \linewidth} \epsfig{file=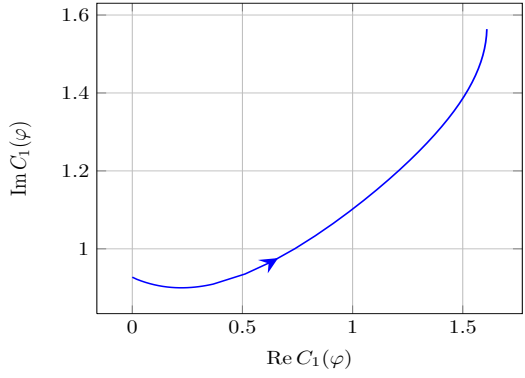, width=7cm,height=5cm}
\captionsetup{width=7cm}
\caption{Path of the system parameter $C_1$. Note that $C_1$ undergoes
only small changes when compared with those seen in $C_2$.\\}
\label{fig:mmpC1}
\end{subfigure}
\begin{subfigure}[b]{.5\linewidth} \centering \epsfig{file=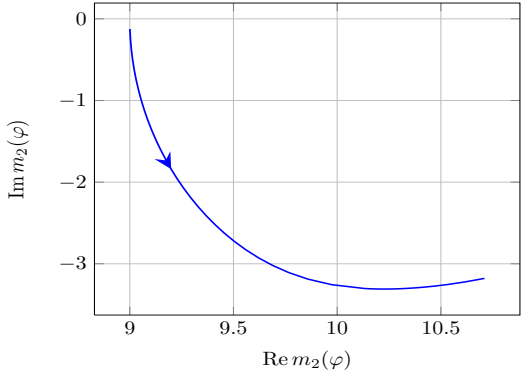,width=7cm,height=5cm}
\captionsetup{width=7cm}
\caption{Path of the system parameter $m_2$. Note that the
absolute value of $m_2$ remains large throughout the
continuation.}
\label{fig:mmpm2}\centering
\end{subfigure}%
\begin{subfigure}[b]{.5 \linewidth} \epsfig{file=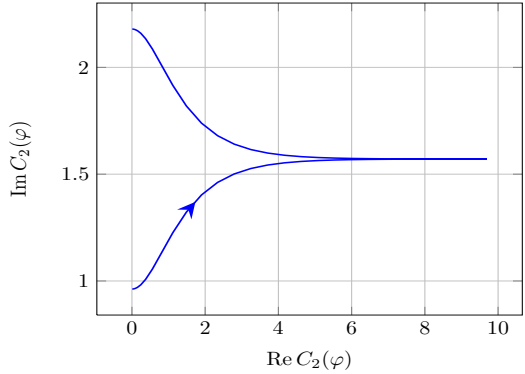, width=7cm,height=5cm}
\captionsetup{width=7cm}
\caption{Path of the system parameter $C_2$. Note that $C_2$ is very strongly
peaked near $\varphi = \pi/2$.\\}
\label{fig:mmpC2}
\end{subfigure}
\caption{In order to describe the path $P_{--+}$ we vary the system parameters
$m_1,m_2$ and $C_1,C_2$ along the curves shown above. As start values
we choose $m_1(0)= 10, m_2(0)= 9$ and $C_1(0) \sim .93i, C_2(0)=.96i$.
These values determine the values of the kinematic variables $w_i$. The
qualitative features of the curves do not depend on the precise values of the
initial conditions. Note that the curves for $m_2$ and $C_2$ are very similar to
those of $m$ and $C$ found in the case of 6-gluon scattering, see
\cite{Bartels:2010ej}.}
\label{fig:mmpmC}
\end{figure}

Let us now describe all this in a bit more detail for the case of $n=7$
gluons and the region $(--+)$ we depict in figure \ref{fig:paths}a. In
terms of the system parameters $m_s$ and $C_s$, $s=1,2$, the path $P_{--+}$
takes the form shown in figure \ref{fig:mmpmC}. Our findings are described
and commented in the captions \footnote{It should be pointed out that there is shift 
in the indices of the variables: for our path $P_{--+}$ the triplet $u_{a\sigma}$ with $\sigma=1$ appears in our final result. However, due to conventions of labeling the functions $Y_{as}$,
the corresponding parameters $m_s$ and $C_s$ carry the label $s=2$.}. 

\begin{figure}
\begin{subfigure}[t]{.5\linewidth} \centering \epsfig{file=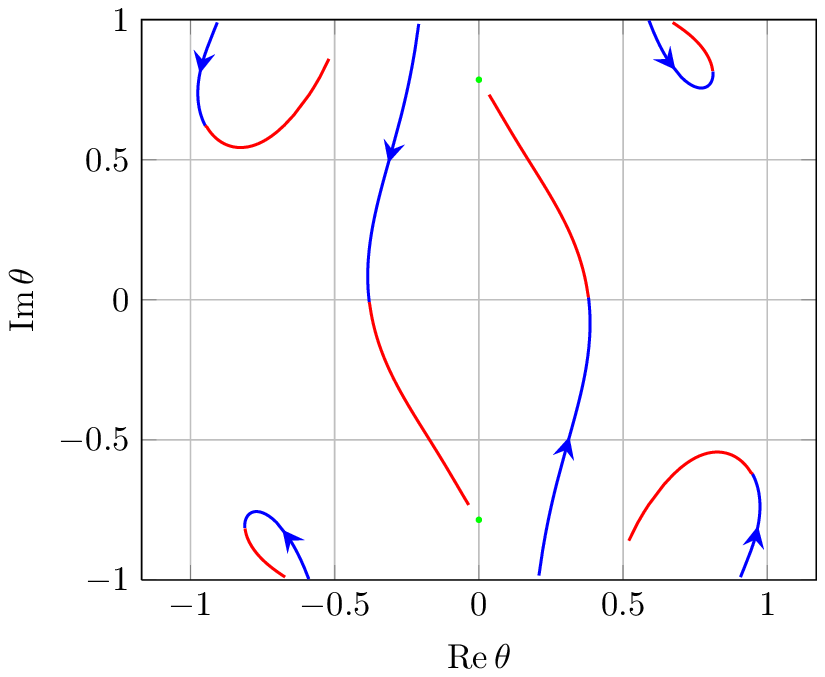,width=7cm,height=5cm}
\captionsetup{width=7cm}
\caption{As we continue along the path $P_{--+}$, two solutions of the
equation $Y_{32}(\theta_*) =-1$ cross the real axis. The green dots indicate the analytically determined endpoints of the crossed solutions.}
\label{fig:mmppolesa}\centering
\end{subfigure}%
\begin{subfigure}[t]{.5 \linewidth} \epsfig{file=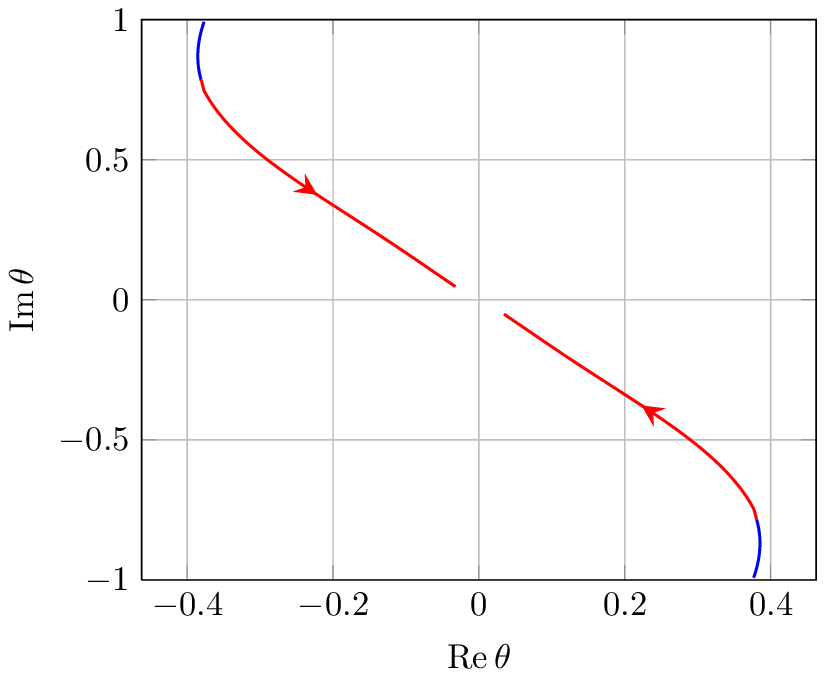, width=7cm,height=5cm}
\captionsetup{width=7cm}
\caption{As we continue along the path $P_{--+}$ two solutions of
$Y_{22}(\theta_*)=-1$ approach the real axis towards the end of the continuation.}
\label{fig:mmppolesb}
\end{subfigure}
\caption{During the continuation along $P_{--+}$ some solutions of
$Y_{as}(\theta_\ast) = -1$ approach or cross the real line. These
are shown in the plots. We change the color of the plot once the pair of solutions crosses the real axis. Solutions of $Y_{a1}(\theta_\ast)=-1$ move
very little and stay away from the real axis. This is related to the
small changes we see in the parameter $C_1$, see figure
\ref{fig:mmpC1}. The pattern of pole crossings is very similar to
that found for 6-gluon scattering, see \cite{Bartels:2010ej}.}
\label{fig:mmppoles}
\end{figure}

\begin{figure}
\begin{subfigure}[b]{.5\linewidth} \centering \epsfig{file=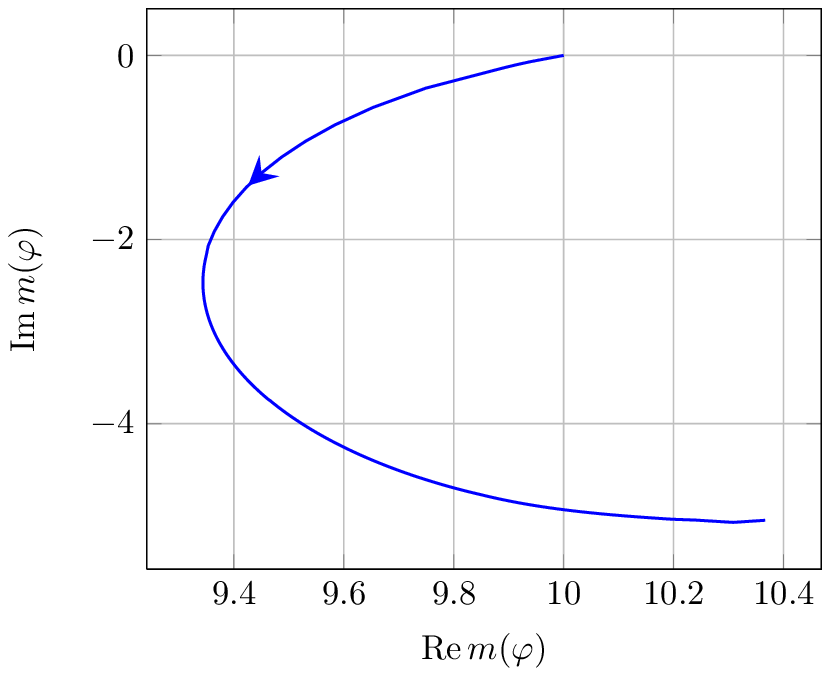,width=7cm,height=5cm}
\captionsetup{width=7cm}
\caption{Path of the system parameter $m=m_1=m_2$. Note that the absolute
value of $m$ remains large throughout the continuation.\\  }
\label{fig:mmmm}\centering
\end{subfigure}%
\begin{subfigure}[b]{.5 \linewidth} \epsfig{file=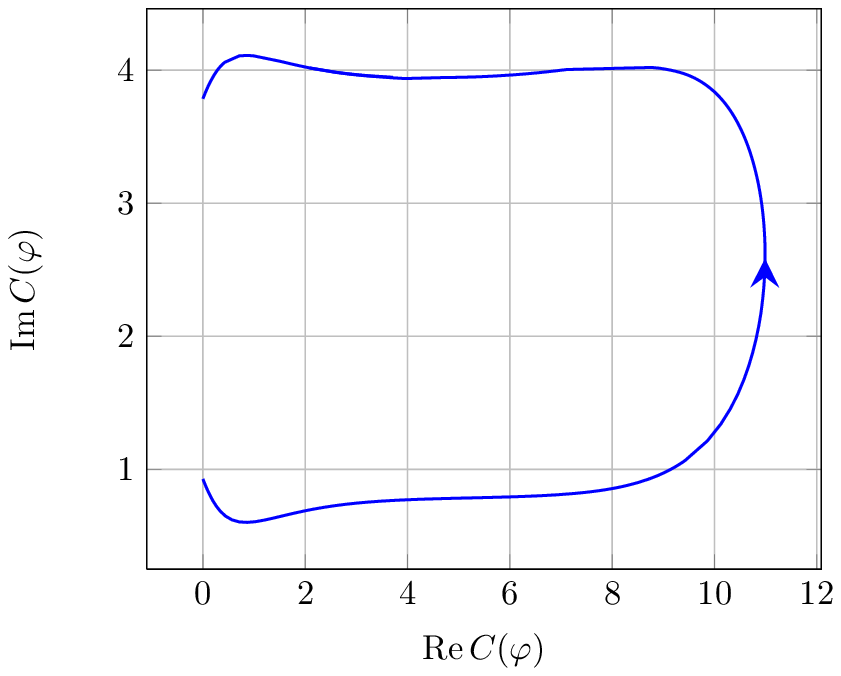, width=7cm,height=5cm}
\captionsetup{width=7cm}
\caption{Path of the system parameter $C=C_1 = -C_2$. Note
that $C$ becomes large (of the order of the mass) again, but
it is not strongly peaked.}
\label{fig:mmmC}
\end{subfigure}
\caption{In order to describe the path $P_{---}$ we vary the system parameters
$m_s$ and $C_s$ along the curves shown in this figure. For these figures we have
chosen symmetric initial conditions $m(0) = m_1(0) = m_2(0) = 10$ and $C(0) =
C_1(0) = - C_2(0) \sim .93i$. For other (non-symmetric) initial conditions the qualitative features of these curves remain the same. Initial conditions must
be varied in order to explore the whole range of dynamical variables $w_1$ and
$w_2$.}
\label{fig:mmmmC}
\end{figure}

\begin{figure}
\begin{subfigure}[b]{.5\linewidth} \centering \epsfig{file=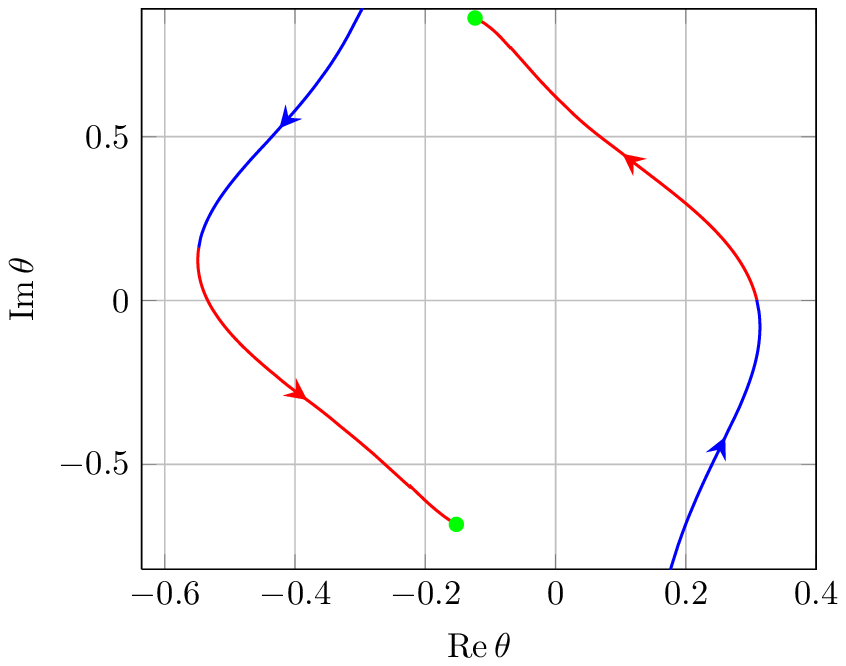,width=7cm,height=5cm}
\captionsetup{width=7cm}
\caption{As we continue along the path $P_{---}$, two of
the solutions to $Y_{12}(\theta_*) = -1$ cross the real
axis. Note that the crossing happens at two different
values of the continuation parameter $\varphi$. The
first pole crossing is correlated with the change of
colors from blue to red.\\}
\label{fig:mmmpoles1}\centering
\end{subfigure}%
\begin{subfigure}[b]{.5 \linewidth} \epsfig{file=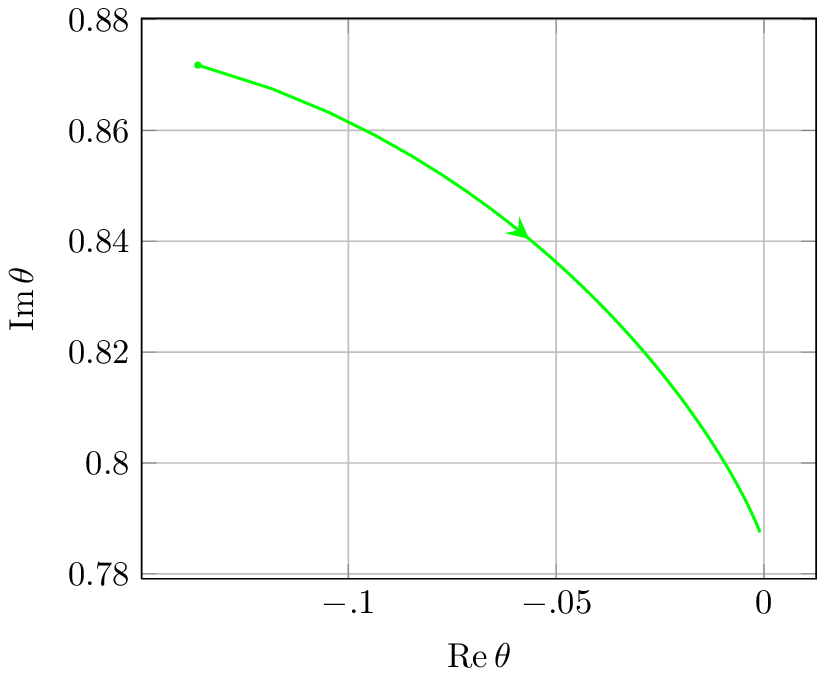,
width=7cm,height=5cm}
\captionsetup{width=7cm}
\caption{The end-point of the right curve in figure \ref{fig:mmmpoles1}
depends on the initial value of $m = m_1 = m_2$. Our figure shows its
position as a function of $m$ in a range between $m=10$ and $m=2000$.
Note that for very large values of $m$, our numerical results approaches
the analytic value $\theta_* = i\pi/4$.}
\label{fig:mmmpoles2}
\end{subfigure}
\caption{Four solutions of the equations $Y_{as} (\theta_*) = -1$ cross the
real axis while we continue along $P_{---}$. We have displayed the relevant
solutions for $Y_{12}$. The corresponding figure for $Y_{31}$ looks very
similar. For all other $Y$ functions, the solutions do not get close to
the real axis. The final position of the solutions at $\varphi = \pi$
determines the free energy. Note that this value is approached very
slowly as we crank up the mass parameter $m$.}
\label{fig:mmmpoles}
\end{figure}

Once these curves are known, we can watch the position of the solutions to $Y(\theta_*)=-1$ to see whether they cross the real axis. In the case of
7 external gluons, there are actually six $Y$-functions which we label by
$Y_{as}$ with $a=1,2,3$ and $s=1,2$. Figure \ref{fig:mmppoles} displays how
the solutions of $Y_{as}(\theta_*) = -1$ move as we vary the system parameters
along the paths of figure \ref{fig:mmpmC}.
We see that a pair of solutions cross the
real axis. The excitations that are created in this process possess
non-vanishing energy which can be computed analytically and contributes
to the following final expression for the remainder function\footnote{Note that contrary to our previous publications we define the remainder function at strong coupling including the factor $-\frac{\sqrt{\lambda}}{2\pi}$ for better comparison with the weak coupling results.}
$$
R^{\infty\, \text{MRL}}_{7,--+} + i\pi \delta_{7,--+}=
{\cal R}^{\infty}(u_{a1}) \ , $$
where
\begin{equation}
  {\cal R}^{\infty}(u_1,u_2,u_3) = \frac{\sqrt{\lambda}}{2\pi}e_2 \ln (1-u_1) +
  \frac{\sqrt{\lambda}}{2\pi}e_2(\frac12 \ln |w|^2 - \ln |1+w|^2)
\end{equation}
and  $e_2 = - \sqrt{2} +\frac{1}{2} \ln (3+2 \sqrt 2)$. This function
is the same that was found previously in the investigation of 6-gluon
scattering amplitudes \cite{Bartels:2013dja}. Note however, that the various
individual contributions to the remainder function must conspire in order to
produce this answer. Of course, one may carry out the same analysis for the path
$P_{+--}$ in order to find
$$
 R^{\infty\, \text{MRL}}_{7,+--}+ i\pi \delta_{7,+--}  =
 {\cal R}^{\infty}(u_{a2})\ \ .
$$
Let us now turn to the most interesting case, namely the path $P_{---}$.
The relevant curves for the mass parameters $m_s$ and the chemical potentials
$C_s$ are shown in figure \ref{fig:mmmmC}. For simplicity we have chosen
symmetric initial conditions $m_1 = m_2 = m$ and $C_1 = -C_2= C$. Since the
path $P_{---}$ respects this symmetry, the equalities remain true for all
values of the continuation parameter $\varphi$.
When we drive the system parameters along these paths, four solutions of $Y_{as}(\theta_\ast) = -1$ cross the real axis. For $Y_{12}$ their paths are shown
in figure \ref{fig:mmmpoles1}. The other two poles are contributed by $Y_{31}$. The
numerical work underlying these plots turned out to be much more intricate than for
the hexagon. In particular, it is quite challenging to determine the positions of
$\theta_\ast$ at the end of the continuation from the numerics. It turns out that
for very large masses $m_s$ the position of these solutions is $\theta_\ast = \pm i \pi/4$, independently of the precise values of the remaining parameters. But the
approach to these final positions requires extremely high masses, see figure \ref{fig:mmmpoles2}. Fortunately, the positions can once again be found
analytically, as will be explained in our forthcoming paper \cite{Bartels:2014}.
The final formula for the resulting remainder function is
\begin{equation}
 R^{\infty\, \text{MRL}}_{7,---} + i\pi \delta_{7,---}=
{\cal R}^{\infty}(u_{a1}) + {\cal R}^{\infty}(u_{a2})\ .
\end{equation}
As in the previous two cases, the expression for the remainder function in the
strongly coupled theory is built from the same function ${\cal R}^\infty$ that
describes the high energy scattering of $6$ gluons. This is in perfect agreement
with the weak coupling expansion in eq.(\ref{R7---}).

There is one curve left to discuss, namely $P_{-+-}$. In this case it
turns out that none of the solutions of $Y_{as}(\theta_\ast) = -1$ actually
cross the real axis. Hence, no excitation is produced so that the remainder
function remains at
\begin{equation} \label{7pmpm}
 R^{\infty, \text{MRL}}_{7,-+-} + i \pi \delta_{7,-+-} = 0\ ,
 \end{equation}
just as in the Euclidean region, except for the appearance of a phase. The corresponding analysis at weak
coupling shows the presence of the Regge cut in the kinematic region $(-+-)$, but there remains the possibility that 
our path of analytic continuation, $P_{-+-}$, needs to be modified.   
It is certainly important to clarify this issue.

\section{Conclusion and Outlook} 

In this note we have reported the results we obtain for the multi-Regge 
limit of 7-gluon scattering amplitudes in strongly coupled $\N=4$ SYM 
theory. These are obtained through a very general algorithm that can in 
principle produce similar formulas for any number of external gluons. 
The general procedure and the technical details that lead to our results 
for 7 gluons are discussed in a forthcoming publication \cite{Bartels:2014}.

There are several interesting directions to pursue. As we have mentioned 
already, the choice of the path $P_{-+-}$ should be studied carefully, in order to see whether results
in the weakly coupled theory  are compatible with what 
we found for strong coupling. Let us also note that all our analysis is 
performed for finite values of the parameters $w_\sigma$. When we send one 
or several of these parameters to zero, some key features of the various
plots we produced numerically could change. So, it may not be correct to 
simply take our expressions for the remainder function and evaluate them 
in the collinear limit $w_\sigma = 0$. The issue certainly requires further 
attention, in particular in view of the close connection between the Wilson 
loop OPE, see \cite{Basso:2013vsa, Basso:2013aha, Basso:2014koa} and references therein, and the collinear 
expansion of the multi-Regge results for 6-gluon amplitudes in the weakly 
coupled theory \cite{Hatsuda:2014oza}. 
\medskip 

\noindent
{\bf Acknowledgments:} We would like to thank
Benjamin Basso, Simon Caron-Huot, Lance Dixon, Andrey Kormilitzin,
Jan Kotanski, Lev Lipatov, Lionel Mason and David Skinner for
useful discussions. This work was supported by the SFB 676. The
research leading to these results has also received funding from
the People Programme (Marie Curie Actions) of the European Union's
Seventh Framework Programme FP7/2007-2013/ under REA Grant Agreement
No 317089 (GATIS).

\end{document}